\documentclass[usenatbib, referee]{mn2e}
\usepackage{graphicx}
\usepackage{epsfig}
\usepackage{color}

\title[MSPs with long orbital periods]
{Formation of millisecond pulsars with long orbital periods by accretion-induced collapse of white-dwarfs}
\author[Wang, Liu \& Chen]
{Bo Wang,$^{\rm 1}$\thanks{E-mail:wangbo@ynao.ac.cn} 
Dongdong Liu$^{\rm 1}$\thanks{E-mail:liudongdong@ynao.ac.cn} 
and Hailiang Chen$^{\rm 1}$\thanks{E-mail:chenhl@ynao.ac.cn}  \\ 
$^1$Yunnan Observatories, Chinese Academy of Sciences, Kunming 650216, China}
\begin{document}
\date{Accepted. Received}
\pagerange{\pageref{firstpage}--\pageref{lastpage}} \pubyear{2022}
\maketitle

\label{firstpage}

\begin{abstract}

Accretion-induced collapse (AIC) of massive white-dwarfs (WDs) 
has been proposed as an  important way for the formation of neutron star (NS) systems.  
An oxygen-neon (ONe) WD that accretes H-rich material from a red-giant (RG) star may experience the AIC process, 
eventually producing millisecond pulsars (MSPs), known as the RG donor channel. 
Previous studies indicate that this channel can only account for MSPs with orbital periods $>500\,\rm d$. 
It is worth noting that some more MSPs with wide orbits ($60-500\,\rm d$) have been detected by recent observations, but their
origin is still highly uncertain.
In the present work, by employing an adiabatic power-law assumptions for the mass-transfer  process, 
we performed a large number of complete binary evolution calculations for the formation of MSPs 
through the RG donor channel in a systematic way. 
We found that this channel can contribute to the observed MSPs with orbital periods in the range of $50-1200\,{\rm d}$, and
almost all the observed MSPs with wide orbits can be covered 
by this channel in the WD companion mass versus orbital period diagram.
The present work indicates that the AIC process provides a viable way to form  MSPs with wide orbits.

\end{abstract}


\section{Introduction} 

It is generally believed that neutron star
(NS) systems can be formed via three ways, i.e.  core-collapse supernovae of massive stars, 
electron-capture supernovae of intermediate-mass stars, and accretion-induced collapse 
(AIC) of massive white-dwarfs (WDs; see van den Heuvel 2009). 
Oxygen-neon (ONe) WDs  are suggested to collapse into neutron stars (NSs) via electron-capture reactions 
once growing in mass approach the Chandrasekhar limit (${M}_{\rm Ch}$),
called the AIC process  (see, e.g.  Nomoto et al. 1979; Miyaji et al. 1980; Taam \& van den Heuvel 1986). 
It has been predicted that AIC events may be the most likely short-lived and  faint optical transients, and   
that a small ejecta mass ($\sim10^{-3}-10^{-1}\,{M}_\odot$) is expected during the collapse
(see, e.g. Woosley \& Baron 1992;  Fryer et al. 1999; Dessart et al. 2006).

A lot of indirect evidence  has been proposed to support the AIC process that
can be used to resolve a variety of troublesome NS systems in observations (for a recent review see Wang \& Liu 2020).
Especially, the AIC process could lead to the formation of newborn NSs with small kicks, and thus it can be used to  reproduce obviously
young  NSs  in some globular clusters; these young NSs cannot be produced 
through the classic core-collapse supernova channel (see Boyles et al. 2011). 
Meanwhile, the AIC process may help to explain the observed discrepancy 
between the large rate of millisecond pulsars (MSPs) and 
the small rate of their progenitor systems (i.e. low-mass X-ray binaries; LMXBs) in the Galaxy
(see, e.g. Kulkarni \& Narayan 1988; 
Bailyn \& Grindlay 1990; Hurley et al. 2010; Tauris et al. 2013).\footnote{MSPs 
are defined as a type of radio pulsars with extremely short spin periods (usually less than 40\,ms),  most of which
are observed in binary systems with WD companions  (see Lorimer 2008).}
In addition, the AIC process has been proposed as  possible production sources of some important events, such as 
gravitational wave radiation (see  Abdikamalov et al. 2010),   elements of r-process nucleosynthesis (e.g.
Wheeler, Cowan \& Hillebrandt 1998; Fryer et al. 1999), and ultrahigh-energy cosmic rays 
(e.g. Usov 1992; Dar et al. 1992; Piro \& Kollmeier 2016), etc.

Although the proposal of the AIC process dates to 40 years ago as a final fate of massive ONe WDs in stellar evolution,
there is no direct detection for such events to date.
Up to now, two families of progenitor models for  AIC events have been proposed, i.e.
the single-degenerate model and the double-degenerate model (for a recent review see Wang \& Liu 2020).
In the single-degenerate model,
an ONe WD grows in mass by accreting H-/He-rich material from a
non-degenerate companion, in which the mass donor  could be a 
main-sequence star (the MS donor channel), 
a red-giant star (the  RG donor channel), or  a He star (the  He star donor channel).
An AIC event may be formed
once the ONe WD increases its mass approach ${M}_{\rm Ch}$ 
(see Nomoto \& Kondo 1991;  Ivanova \& Taam 2004; 
Tauris et al. 2013; Brooks et al. 2017; Wang 2018; Liu et al. 2018; Ruiter et al. 2019). 
Recent studies  indicate that carbon-oxygen (CO) WD+He star binaries may also produce NS systems via AIC  
when off-centre carbon burning happens on the CO WD  (see Brooks et al. 2016; Wang, Podsiadlowski \& Han 2017).
In the double-degenerate model, an AIC event originates from the merging of double WDs 
with a total mass larger than ${M}_{\rm Ch}$; the merging of double WDs was caused 
due to orbit shrinking resulting from gravitational wave radiation, 
called the merger-induced collapse (e.g. Nomoto \& Iben 1985; Saio \& Nomoto 1985;  Liu \& Wang 2020).

The merger-induced collapse  mainly produces isolated NSs, whereas the single-degenerate model  forms NS binaries.
The post-AIC systems in the  single-degenerate model may evolve to some different  kinds of NS binaries, 
mainly depending on the chemical composition of the mass donors.
It has been suggested that ONe WDs with MS donors
could produce fully recycled  MSPs with orbital periods $\sim10-60$\,d (see Tauris et al. 2013; see also Hurley et al. 2010).
Liu et al. (2018) recently investigated the He star donor channel for the production of intermediate-mass binary pulsars (IMBPs),
and they suggested that this channel may reproduce the properties of most of the observed IMBPs with orbital periods $<10$\,d. 
Especially, the  He star donor channel can account for the observed parameters of 
PSR J1802$-$2124  that is one of the two well observed IMBPs (see Liu et al. 2018).

Tauris et al. (2013) studied the binary evolution calculations of ONe WD+RG systems 
that experience AIC and then be recycled to produce MSPs, 
where they only considered the case with the initial ONe WD mass of $1.2\,\rm M_{\odot}$. They argued that 
ONe WDs with RG donors  could potentially be identified as
LMXBs, eventually forming young  MSPs with more mildly recycled spins and long orbital periods ($>500\,\rm d$).
We note that  some more MSPs with orbital periods between $60\,\rm d$ and $500\,\rm d$  
have been detected by recent observations (see, e.g. Bhattacharyya et al. 2019; Bondonneau et al. 2020; 
Deneva et al. 2021), but their origin is still highly unclear.
Wang (2018) recently investigated the binary evolution of semidetached ONe WD+RG systems for producing AIC events, 
but they did not consider the evolution of post-AIC systems for the formation of MSPs. 
In this work, we mainly explore whether the RG donor channel can account for  the formation of these MSPs with wide orbits.

The purpose of  this article is to investigate the formation of MSPs 
through the RG donor channel comprehensively using an integrated mass-transfer prescription. We found that 
this channel can contribute to MSPs with orbital periods in the range of $50-1200\,{\rm d}$.
In Section 2, we introduce the numerical methods  and physical assumptions of binary evolution computations of ONe WD+RG systems,
including the evolution of pre-/post-AIC systems. The corresponding results of binary evolution are given  in Section 3.
Finally, a discussion is given in Section 4 and a summary in Section 5.

\section{Numerical methods and assumptions}

In the present work, the numerical models consist of two parts: i) evolution of 
ONe WD+RG systems prior to AIC (i.e. pre-AIC evolution; see Sect. 2.1) 
and ii) evolution of  NS+RG systems after AIC (i.e. post-AIC evolution; see Sect. 2.2).
Both parts are calculated with the Eggleton stellar evolution code (see Eggleton 1971, 1972, 1973), in which
the input physics was updated over the past five decades (see Han, Podsiadlowski \& Eggleton 1994; Pols et al.\ 1995, 1998;
Eggleton \& Kiseleva-Eggleton 2002). We set the convective overshooting parameter ($\delta_{\rm OV}$) to be 0.12, 
and the ratio of mixing length to local pressure scale height to be 2.0, 
approximately being equivalent to an overshooting length of $\sim$0.25 pressure scale heights (see Pols et al.\ 1997). We adopt
a classic composition of Population I  for the initial MS models with metallicity $Z=0.02$, H
abundance $X=0.70$, and He abundance $Y=0.28$. 
For pre-/post-AIC evolution, we do not calculate the structure of the mass-accreting star (i.e. ONe WD or NS) and set it to be a point mass. 

During Roche-lobe overflow (RLOF),
the mass-transfer process for the evolution of pre-/post-AIC systems is computed by an integrated mass-transfer prescription provided by Ge et al. (2010), 
which is more reasonable for semidetached binaries with giant donors (for details see Liu et al. 2019, and references therein).
The mass-transfer rate can be expressed as
\begin{equation}
\dot{M}_{\rm 2}=-\frac{2\pi R_{\rm L}^{\rm 3}}{GM_{\rm 2}}f(q)
\int_{\rm \phi_{\rm L}}^{\rm \phi_{\rm s}}\Gamma^{\rm 1/2}(\frac{2}{\Gamma+1})^{\rm \frac{\Gamma+1}{2(\Gamma-1)}}(\rho P)^{\rm 1/2} \rm d\phi,
\end{equation}
where $G$ is the gravitational constant, $R_{\rm L}$ is the effective radius of the Roche-lobe, 
$M_{\rm 2}$ is the donor mass, $P$ is the local gas pressure, $\rho$ is the local gas density, 
$\Gamma$ is the adiabatic index, $\phi_{\rm s}$ is the stellar surface potential energy,
and $\phi_{\rm L}$ is the Roche-lobe potential energy
 (see Appendix of Ge et al. 2010). 
For more discussions on the mass-transfer process, see Sect. 4.1.

\subsection{Pre-AIC evolution}

For pre-AIC evolution, the ONe WD starts to accrete
H-rich material from its RG companion once RLOF occurs. 
The accreted H-rich material is processed into He, and 
then the He is burned into C and O, resulting in the mass growth of the ONe WD.  
When the ONe WD increases its mass to $1.38\,\rm M_{\odot}$  (a critical mass limit 
for non-rotating ONe WDs;  see Wu \& Wang 2018), 
we suppose that the ONe WD will undergo AIC process and collapse into a NS; 
the maximum stable mass for 
rotating ONe WDs is likely to be above the standard ${M}_{\rm Ch}$ 
(see, e.g. Yoon \& Langer 2005; Chen \& Li 2009; Wang et al. 2014; Freire \& Tauris 2014).
The basic initial setup  and input physics for pre-AIC evolution are similar to those in Wang (2018).

During the mass-transfer process,  we define the WD mass-growth rate  ($\dot {M}_{\rm WD}$) as
\begin{equation}
\dot{M}_{\rm WD}=\eta_{\rm H}\eta_{\rm He}\dot{M}_{\rm 2},
\end{equation}
where $\eta_{\rm H}$ is the mass-retention efficiency of H-shell burning from Wang, Li \& Han (2010), 
and $\eta_{\rm He}$ is the mass-retention efficiency of He-shell flashes from Kato \& Hachisu (2004).
If $\dot{M}_{\rm 2}$ is lower than the minimum accretion rate 
for stable H-shell burning ($\dot{M}_{\rm st}$),  the accreted H-rich material will undergo H-flashes on the surface of the WD. 
When $\dot{M}_{\rm 2}$ is larger than a critical rate  ($\dot {M}_{\rm cr}$) for stable H-shell burning (see Nomoto 1982), we suppose that 
the accreted H-rich material  burns into He at the rate of $\dot {M}_{\rm cr}$ in a stable way, 
but the rest will be blown away via the optically thick wind (see Hachisu, Kato \&  Nomoto 1996). 
If $\dot{M}_{\rm 2}$ is lower than $\dot {M}_{\rm cr}$ but higher than $\dot{M}_{\rm st}$, 
we suppose that the H burns in a stable way and no mass is lost from the system. 
For more studies on the mass-accretion onto WDs, see, e.g. 
Langer et al. (2000), Yoon, Langer \& Scheithauer (2004), Han \& Podsiadlowski (2006), Ruiter, Belczynski \& Fryer (2009),  
Toonen, Voss \& Knigge (2014),  Meng \& Podsiadlowski (2017), Soker (2018), Chen et al. (2019), etc.

\subsection{Post-AIC evolution}

We suppose that AIC happens once the ONe WD grows in mass to  $1.38\,\rm M_{\odot}$.
During the AIC process, we assume that the ONe WD loses $0.13\,\rm M_{\odot}$ material that 
is converted into the released gravitational binding energy, finally collapses into
 a NS with the gravitational mass of $1.25\,\rm M_{\odot}$   (see Ablimit \& Li 2015). 
Owing to the sudden mass-loss from the system, the orbital separation of the binary becomes wider. 
In the present work, we suppose that the binary orbit is re-circularized after AIC,
and the orbital separation ($a$) after AIC is calculated   on the basis of 
angular momentum conservation, written as
\begin{equation}
{a}=a_{\rm 0}\frac{M_{\rm WD}+M_{\rm 2}}{M_{\rm NS}+M_{\rm 2}},
\end{equation}
in which  $a_{\rm 0}$ and $M_{\rm WD}$ are the orbital separation of binary and  the mass of the ONe WD at the moment of AIC, 
$M_{\rm NS}$ and $M_{\rm 2}$ are the masses of the newborn NS and
 the RG star just after AIC, respectively (see Verbunt, Wijers \& Burn 1990). 
It is generally believed that the AIC  process possibly leads to a small kick or no kick (e.g. Boyles et al. 2011).
Previous studies found that  a kick velocity with a dispersion of $50\,\rm km/s$ has 
no big influence on the final results (for details see Tauris et al. 2013). 
Accordingly, we ignore the influence of the kick velocity on the newborn NS in the present work.

After the AIC process, the RG star may fill its Roche-lobe again, and transfer H-rich material
and angular momentum  onto the newborn NS (known as the recycling process),
during which  post-AIC systems with RG donors could potentially be identified as
LMXBs and the resulting MSPs.
The prescription of Tauris et al. (2013) is adopted  to calculate the NS mass-growth rate, written as
\begin{equation}
\dot{M}_{\rm NS}=(|\dot{M}_{\rm 2}|-\max[|\dot{M}_{\rm 2}|-\dot{M}_{\rm Edd},0])\cdot k_{\rm def}\cdot e_{\rm acc},
\end{equation}
where $k_{\rm def}$ is the ratio between the gravitational mass and the rest mass of the accreted material, 
$e_{\rm acc}$ is the mass fraction between the transferred material from the RG star and the remains on the NS, 
and  $\dot{M}_{\rm Edd}$ is the Eddington accretion rate.
Here, $\dot{M}_{\rm Edd}$ can be written as
\begin{equation}
\dot{M}_{\rm Edd}=2.3\times 10^{\rm -8}\cdot M^{\rm -1/3}_{\rm NS}\cdot \frac{2}{1+X}\,\rm M_{\odot}\,\rm yr^{\rm -1},
\end{equation}
in which $X$ is the H mass fraction of the accreted material.

In this work, we combine  $k_{\rm def}$ and $e_{\rm acc}$ into a free parameter $k_{\rm def}\cdot e_{\rm acc}$ (i.e.  retention efficiency), 
and set $k_{\rm def}\cdot e_{\rm acc}=0.35$ based on the increasing facts of the inefficient 
accretion for LMXBs  (see, e.g. Jacoby et al. 2005; Antoniadis et al. 2012; Ablimit \& Li 2015). 
According to Eq.\,(4),  the values of
$\dot{M}_{\rm NS}$ are:
 \begin{equation}
\dot{M}_{\rm NS}=\left\{
 \begin{array}{ll}
0.35\dot{M}_{\rm Edd}, & |\dot{M}_{\rm 2}|\geq \dot{M}_{\rm Edd},\\
0.35\dot{M}_{\rm 2}, & |\dot{M}_{\rm 2}|< \dot{M}_{\rm Edd}.
\end{array}\right.
\end{equation}
The post-AIC evolution with RG donors is  similar to  that of normal LMXB evolution.
The only difference is that the RG donor in this work has already lost some of its material during the pre-AIC evolution.

For the post-AIC system, the accreted material onto the NS ($\Delta M_{\rm NS}$) may recycle the newborn NS that 
will experience spin-up/-down processes.
We employ the prescription of Tauris, Langer \& Kramer (2012) to calculate the minimum spin period
of the recycled NS before spin-down process, written as
\begin{equation}
P_{\rm spin}^{\rm min}\sim 0.34\times(\Delta M_{\rm NS}/\rm M_{\odot})^{\rm -3/4},
\end{equation}
where $P_{\rm spin}^{\rm min}$ is in units of ms.
It is worth noting that this prescription relies on two assumptions, as follows: 
(i) the initial spin velocity of the newborn NS is negligible before 
the mass-accretion,  
and (ii)  the gravitational binding energy of the accreted material onto the newborn NS is negligible.

\section{Numerical results}

We carried out a large number of complete binary evolution calculations of 
ONe WD+RG systems for the formation of MSPs.
Table 1 lists the main evolutionary properties of some typical ONe WD+RG systems that can evolve into MSPs.
In this table,  we first explored the effect of different initial orbital periods  on the final results (see sets $1-10$), 
and then the effect of different initial companion masses  (see sets $11-14$).

\begin{table*}
\begin{center}
\caption{A summary of selected ONe WD+RG systems that can evolve successfully to AIC events and later form recycled MSPs
with different initial orbital periods and initial masses of the donors, in which we set $M_{\rm WD}^{\rm i}=1.2\,\rm M_{\odot}$.
The columns (from left to right):  the initial mass of the donor  and the initial orbital period; 
the stellar age of the donor at the onset of RLOF;
the donor mass  and  the orbital period just after AIC; the time-scale that the binary appears as a LMXB; 
the final NS mass, the final donor mass, and the final orbital period; and the minimum spin period based on Eq. (7).}
\begin{tabular}{ccccccccccccccc}
\hline\hline
Set & $M_2^{\rm i}$ & $\log P_{\rm orb}^{\rm i}$ &  $t_{\rm RLOF}$ & $M_2^{\rm AIC}$ & $\log P_{\rm orb}^{\rm AIC}$ 
& $\bigtriangleup t_{\rm LMXB}$ &  $M_{\rm NS}^{\rm f}$ & $M_2^{\rm f}$ & $\log P_{\rm orb}^{\rm f}$  &  $P_{\rm spin}^{\rm min}$\\
 &  ($\rm M_{\odot}$) & (d)  & (Gyr)   &  ($\rm M_{\odot}$)  &  (d) &  (Myr)  & ($\rm M_{\odot}$) & ($\rm M_{\odot}$)  & (d) & (ms)\\
\hline
 1&  1.4 & $0.5$ &  3.469  & 1.0386 & 0.6257  &  180&1.5087 &0.2995  &1.7483 &   0.94  \\
  2&  1.4 & $0.6$ &  3.496 & 1.0217 & 0.7358 &  145&1.5017  &0.3062 &1.8248 &   0.96  \\
  3&  1.4 & $0.8$ &  3.537 & 0.9890 & 0.9574 &  94 &1.4857   &0.3203 &1.9756 &    1.01  \\
  4& 1.4 & $1.0$ &  3.569 & 0.9829 & 1.1617 &  45  &1.4705   &0.3353 &2.1252 &   1.06  \\
  5& 1.4 & $1.2$ &  3.601 &0.9722 &1.3711  &   29 &1.4403    &0.3520 &2.2758 &    1.18  \\
  6& 1.4 & $1.4$ &  3.613  &0.9572 &1.5837 &   19  &1.3984   &0.3712 &2.4281 &    1.43  \\
  7& 1.4 & $1.6$ &   3.621  & 0.9443 &1.7946 &   13 &1.3538  &0.3931 &2.5797 &    1.88 \\
  8& 1.4 & $1.8$ &   3.627  &0.9530 &1.9902 &     9  & 1.3217 &0.4179 &2.7247 &    2.46 \\
  9& 1.4 & $2.0$ &   3.631  &0.9648 &2.1836 &     6 &  1.3003 &0.4464 &2.8623 &    3.20 \\
  10& 1.4& $2.2$ &   3.634  &0.9800 &2.3740 &      4&1.2851  &0.4705 &2.9918 &     4.28  \\
     &      &            &              &            &            &        &             &            &            &       \\
11&  1.1&   1.8    &   8.433  &0.8693 &1.9418 & 10   &1.3311  &0.4026 &2.6460 &    2.25 \\
12&1.3 &    1.8    &   4.699   &0.9509 &1.9632 & 9  &1.3263   &0.4138 &2.7057 &    2.37 \\
13&1.5 &    1.8    &    2.887  &0.9258 &2.0294 &  8 &1.3176   &0.4213 &2.7362 &    2.60 \\
14&1.7 &    1.8    &     1.950 &0.8217 &2.1297 &  7 &1.3058   &0.4257 &2.7374 &    2.98 \\
\hline
\end{tabular}
\end{center}
\end{table*}

\subsection{A typical example for binary evolution}

\begin{figure*}
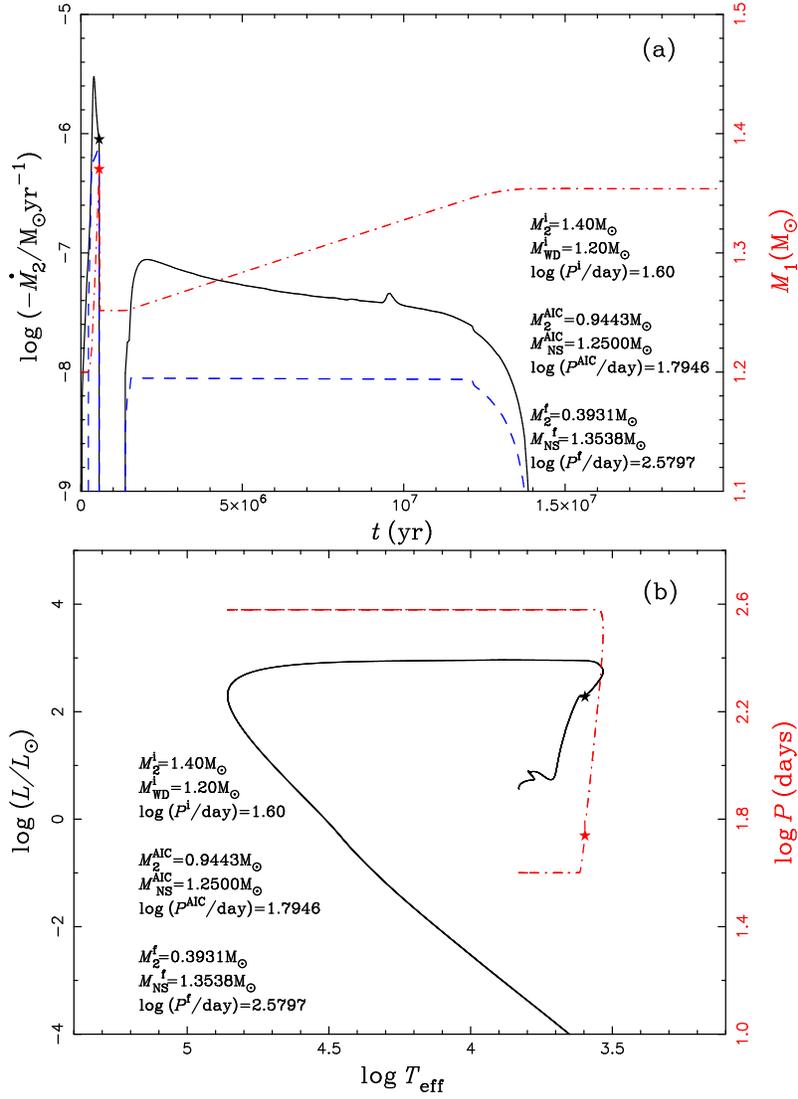

\epsfig{file=f1a.eps,angle=270,width=10.5cm}\ \
\epsfig{file=f1b.eps,angle=270,width=10.5cm}
  \caption{A typical example for the evolution of an ONe WD$+$RG system until the formation of a recycled MSP, where 
  ($M_2^{\rm i}$, $M_{\rm WD}^{\rm i}$, $\log (P_{\rm orb}^{\rm i}/{\rm d})$)
$=$ (1.4, 1.2, 1.6) (see set 7 in Table 1).
Panel (a): the evolution of the $\dot{M}_{\rm 2}$ (black solid curve), $\dot{M}_{\rm WD}$ (or $\dot{M}_{\rm NS}$;  blue dashed curve) 
and $M_{\rm WD}$ before AIC (or $M_{\rm NS}$ after AIC; red  dash-dotted curve) as a function of time for the binary evolution computations. 
Panel (b): the luminosity of the mass donor (black solid curve) and the binary orbital
  period (red dash-dotted curve) as a function of effective temperature.
  The filled stars in panels (a) and (b) represent the position where AIC occurs.
 The initial binary parameters of ONe WD$+$RG system, the binary parameters of 
 NS$+$RG systems just after AIC, and the final binary parameters of  
 the eventually formed NS$+$WD systems are given in these two panels.}
\end{figure*}

Fig.\,1 shows a typical example of the evolution of an ONe WD+RG system that 
experiences AIC process and eventually forms a MSP  (see set 7 in Table 1).
The initial parameters for this binary are
($M_2^{\rm i}$,  $M_{\rm WD}^{\rm i}$, $\log (P_{\rm orb}^{\rm i}/{\rm d})$)
$=$ (1.4, 1.2, 1.6), in which $M_2^{\rm i}$, $M_{\rm WD}^{\rm i}$ and $P_{\rm orb}^{\rm i}$ are the initial
masses of the ONe WD and the donor star in solar masses, and the initial orbital period in days, respectively. 
The mass donor  fills its Roche-lobe due to rapid expansion  of itself  when it evolves to the RG stage, resulting in case B mass-transfer. 
The mass-transfer rate $\dot{M}_{\rm 2}$ is larger than the maximum critical accretion rate for stable H-shell burning
 soon after the onset of RLOF, leading to a wind phase. 
During this stage,  part of the transferred material is blown off in the form of the optically
thick wind, and the left  is accumulated onto the surface of the ONe WD at a rate of $\dot {M}_{\rm cr}$. 
After about $5.5\times10^{\rm 5}\,\rm yr$, an AIC event is expected to occur when the ONe WD grows in mass to $1.38\,\rm M_{\odot}$.
At this moment, the mass of the donor is $0.9443\,\rm M_{\odot}$ and the orbital period is $55.5\,\rm d$. 

During the AIC process, the ONe WD will collapse into a NS, 
and an  equivalent mass of $0.13\,\rm M_{\odot}$ material is converted into the released gravitational binding energy.
After the AIC process, the binary turns to have a $1.25\,\rm M_{\odot}$ NS and a $0.9443\,\rm M_{\odot}$ evolved 
RG star with an orbital period of $62.3\,\rm d$. 
At this time, the radius of the RG star  turns to be below its Roche-lobe radius since the binary orbit becomes wider after the extreme mass-loss. 
After about $8.0\times10^{\rm 5}\,\rm yr$, the evolved RG star refills its Roche-lobe and begins to 
transfer H-rich material onto the surface of the newborn NS again, 
resulting in a spin-up process for the NS. 
The mass-transfer rate increases quickly and its value is higher than $\dot{M}_{\rm Edd}$  soon after the RG star refills its Roche-lobe.
In this case, the NS increases its mass slowly at a rate of $0.35\dot{M}_{\rm Edd}$. 
Meanwhile,  the majority of the transferred matter is blown away from the system 
at a rate of ($|\dot{M}_{\rm 2}|-0.35\dot{M}_{\rm Edd}$), driven by the radiation pressure of the NS.
At this stage, the binary appears to be a LMXB, lasting for about $13\,\rm Myr$.
After that, the RG donor exhausts its H-shell and evolves to a He WD. 
The binary eventually evolves into a long orbital MSP consisting of  a $1.3538\,\rm M_{\odot}$ pulsar  
and a $0.3931\,\rm M_{\odot}$ He WD with an orbital period of  about $380\,\rm d$,
in which the minimum spin period  for the pulsar
will  approach about $1.88\,\rm ms$ before the spin-down process.

\subsection{Initial parameters of ONe WD$+$RG systems}

\begin{figure}
\begin{center}
\epsfig{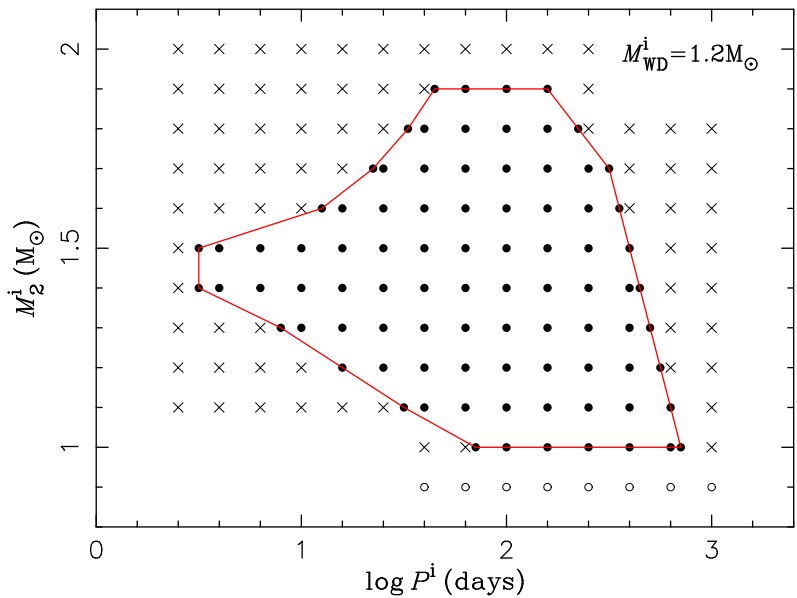} 
\caption{Initial parameter space of ONe WD$+$RG systems that eventually form MSPs 
in the $\log P^{\rm i}-M^{\rm i}_2$ plane  with ${M}^{\rm i}_{\rm WD}=1.2\,\rm M_{\odot}$.
The filled circles denote systems that experience the AIC process, resulting in the formation of MSPs.
The crosses indicate systems that will not experience the AIC process. 
Open circles are those that the mass donors have stellar age larger than the Hubble time when they fill their Roche-lobes. }
\end{center}
\end{figure}

\begin{figure}
\begin{center}
\epsfig{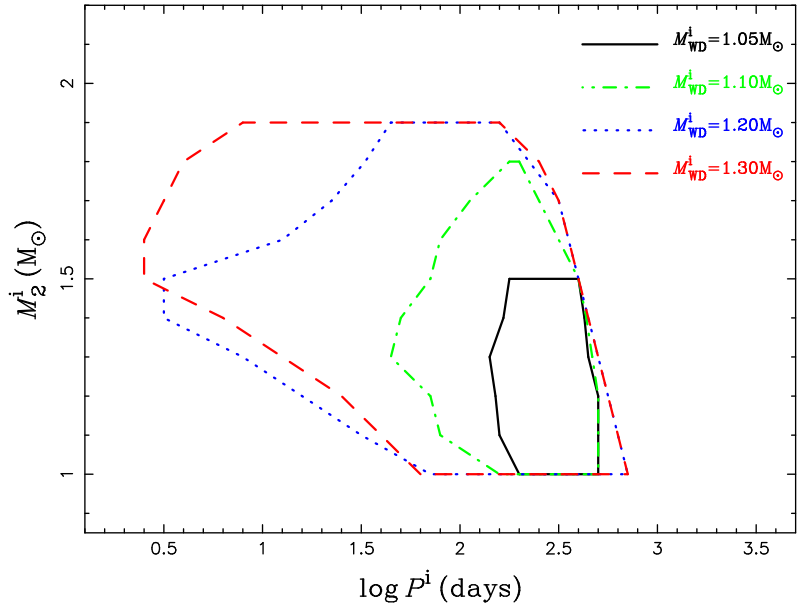}
 \caption{Initial parameter space of ONe WD$+$RG systems that eventually form MSPs 
 in the $\log P^{\rm i}-M^{\rm i}_2$ plane with various initial WD masses. }
  \end{center}
\end{figure}

We performed a large number of detailed binary evolution calculations of ONe WD+RG systems for the production of MSPs, 
thus obtaining a dense binary grid. 
Fig.\,2 shows the initial parameter regions of ONe WD$+$RG systems that eventually form MSPs in 
the $\log P^{\rm i}-M^{\rm i}_2$ plane with ${M}^{\rm i}_{\rm WD}=1.2\,\rm M_{\odot}$, 
where $M^{\rm i}_2$ is the initial mass of the RG star and $P^{\rm i}$ is the initial orbital period  of the  ONe WD+RG system. 
In this figure, the filled circles denote systems that experience the AIC process, eventually producing MSPs.
The crosses denote systems that will not produce AIC events, and the
open circles are those that the mass donors have stellar age larger than the Hubble time when they fill their Roche-lobes.

Fig.\,3 presents the initial parameter regions of ONe WD$+$RG systems that eventually form MSPs in 
the $\log P^{\rm i}-M^{\rm i}_2$ plane for different ${M}^{\rm i}_{\rm WD}$, 
in which we set the minimum mass of ONe WDs to be
$1.05\,\rm M_{\odot}$ (see also Siess 2007).
The boundaries of the initial regions are mainly constrained by the following conditions:
(1)
The upper boundaries are constrained by a high
mass-transfer rate owing to a large mass-ratio between the RG star
and the ONe WD, leading to the formation of a common envelope. 
(2) 
Binaries beyond the right boundaries will undergo a relatively high mass-transfer process as a result of the rapid
expansion of the RG stars, losing too much mass in the form of the optically thick wind. 
(3) 
Binaries beyond the left boundaries will experience strong H-shell flashes due to low mass-transfer rates,
 preventing the WDs from growing in mass to $1.38\,\rm M_{\odot}$. Note that the mass donors with the shortest orbital period for the cases of
 $M^{\rm i}_{\rm WD}=1.20$ and $1.30\,\rm M_{\odot}$ would fill their Roche-lobes at the bottom of the RG stage. 
(4)
Binaries below the lower boundaries have the mass donors with stellar age larger than 
the Hubble time before filling their Roche-lobes. 

In order to produce AIC events, the ONe WD should have RG companions with initial masses of $\sim1.0-1.9\,\rm M_{\odot}$ 
and initial orbital periods of $\sim3-800$\,d.  
The  time-scales that the post-AIC systems appear as LMXBs 
are in the range of $\sim10^{6}-10^{8}\,\rm yr$.
If the initial parameters of an
ONe WD+RG system are located in the
contours of Fig. 3, a MSP is then supposed to be formed. 
In the present work, we set  a typical  Population I composition with solar metallicity for the binary evolution calculations.
If a lower metallicity is adopted, the initial contours for producing AIC events in Fig.\,3 would move to
lower mass donors and shorter orbital periods  (see Tauris 2013; see also Meng et al. 2009; Wang \& Han 2010). 
For more discussions on the initial parameter regions of Fig. 3, see Sect. 4.1.

\subsection{Resulting MSPs}

\begin{table*}
\begin{center}
 \caption{The relevant parameters of 39 observed binary pulsars that have WD companions with 
 long orbital periods ($>$50\,d), 
in which 35 of them are MSPs that have spin periods $<40\,{\rm ms}$. 
 The data was 
 taken from ATNF Pulsar Catalogue in January 2022 
 (see Manchester et al. 2005; see http://www.atnf.csiro.au/research/pulsar/psrcat). 
The median  masses of the WD companions are computed by supposing a typical pulsar mass of $1.35\,\rm M_{\odot}$
and an orbital inclination angle of $60^{\circ}$  for simplicity. 
The error bars of the WD masses correspond to an unknown orbital inclination angle, as follows:
the lower limit marks an inclination angle of $90^{\circ}$, whereas  the upper limit represents a 90\% probability limit. 
If a larger value of the pulsar mass is adopted, the error bars of the WD masses will become 
larger based on the binary mass function.}
   \begin{tabular}{ccccccccc}
\hline \hline
 $\rm No.$ & $\rm Pulsars$ & $P_{\rm spin}$ & $P_{\rm orb}$ & $M_{\rm WD}$\\
                  &                        &               (ms)            &  (d)   &   ($\rm M_{\odot}$) \\
\hline
1  & $\rm J0214+5222$ & 24.58 &  512.0  &  $0.48^{+0.70}_{-0.07}$ \\   
2  &$\rm J0407+1607$ &25.70  &  669.1  &  $0.22^{+0.27}_{-0.03}$\\     
3 & $\rm J0605+3757$ &2.73     &55.7    & $0.32^{+0.24}_{-0.03}$\\       
4 &  $\rm J0614-3329$ &3.15     &53.6     &$0.32^{+0.42}_{-0.05}$\\      
5  & $\rm J1125-5825$ & 3.10  & 76.4     &  $0.31^{+0.40}_{-0.05}$\\     
6   & $\rm J1312+1810$&33.16  &255.8  &   $0.35^{+0.47}_{-0.05}$\\     
7 & $\rm J1342+2822D$ & 5.44   &128.8   &  $0.24^{+0.30}_{-0.04}$\\  
8  & $\rm J1455-3330$ &7.99   &76.2      & $0.30^{+0.38}_{-0.04}$\\      
9  & $\rm J1516-43$&36.02 & 228.4       &   $0.47^{+0.68}_{-0.07}$\\    
10  & $\rm J1529-3828$&8.49   &119.7    &$0.19^{+0.22}_{-0.03}$\\      
11  & $\rm J1536-4948$&3.08  & 62.1    &$ 0.32^{+0.42}_{-0.05}$\\      
12  & $\rm J1623-2631$&11.08 & 191.4  & $0.33^{+0.42}_{-0.05}$\\     
13  & $\rm J1640+2224$&3.16   &175.5   &$0.29^{+0.37}_{-0.04}$\\     
14  & $\rm J1643-1224$&4.62   &147.0   &$0.14^{+0.15}_{-0.02}$\\      
15  & $\rm J1708-3506$&4.51   &149.1   &$ 0.19^{+0.22}_{-0.03}$\\     
16  & $\rm J1711-4322$&102.62 &922.5 & $0.24^{+0.29}_{-0.03}$\\     
17  & $\rm J1713+0747$&4.57   &67.8    &$0.32^{+0.42}_{-0.05}$\\      
18  & $\rm J1748-2446E$&2.20   &60.1   &$0.25^{+0.30}_{-0.04}$\\     
19  & $\rm J1751-2857$&3.91   &110.7   & $ 0.23^{+0.27}_{-0.03}$\\    
20  & $\rm J1801-0857B$&28.96 & 59.8   &$ 0.38^{+0.51}_{-0.06}$\\   
21  & $\rm J1822-0848$&834.84 & 286.8   &$ 0.38^{+0.52}_{-0.06}$\\  
22  & $\rm J1824+1014$&4.07   &82.6     & $0.31^{+0.40}_{-0.05}$\\    
23  & $\rm J1825-0319$ &4.55   &52.6     & $0.21^{+0.24}_{-0.03}$\\    
24  & $\rm J1844+0115$&4.19   &50.6    &$0.16^{+0.18}_{-0.02}$\\      
25  & $\rm J1850+0124$&3.56   &84.9   & $0.29^{+0.36}_{-0.04}$\\     
26  & $\rm J1853+1303$&4.09   &115.7 & $ 0.28^{+0.35}_{-0.04}$\\    
27  & $\rm J1855-1436$&3.59   &61.5   & $ 0.31^{+0.40}_{-0.05}$\\     
28  & $\rm J1910+1256$&4.98   &58.5  & $ 0.22^{+0.27}_{-0.03}$\\     
29  & $\rm J1913+0618$&5.03   &67.7  & $ 0.33^{+0.43}_{-0.05}$\\     
30  & $\rm J1930+2441$&5.77   &76.4  & $ 0.27^{+0.33}_{-0.04}$\\     
31  & $\rm J1935+1726$&4.20   &90.8  & $ 0.26^{+0.31}_{-0.04}$\\     
32  & $\rm J1955+2908$&6.13  & 117.3 &$  0.21^{+0.25}_{-0.03}$\\    
33  & $\rm J2016+1948$&64.94 & 635.0 &$ 0.34^{+0.45}_{-0.05}$\\    
34  & $\rm J2019+2425$&3.93  & 76.5   & $0.36^{+0.49}_{-0.06}$\\     
35  & $\rm J2033+1734$&5.95   &56.3   & $0.22^{+0.26}_{-0.03}$\\     
36  & $\rm J2042+0246$&4.53   &77.2   &$ 0.22^{+0.26}_{-0.03}$\\     
37  & $\rm J2204+2700$&84.70 &815.2  &$ 0.42^{+0.58}_{-0.06}$\\    
38  & $\rm J2229+2643$&2.98   &93.0   & $0.14^{+0.16}_{-0.02}$\\      
39  & $\rm J2302+4442$&5.19  & 125.9 & $ 0.34^{+0.45}_{-0.05}$\\     

\hline \label{1}
\end{tabular}
\end{center}
\end{table*}

\begin{figure}
\begin{center}
\epsfig{file=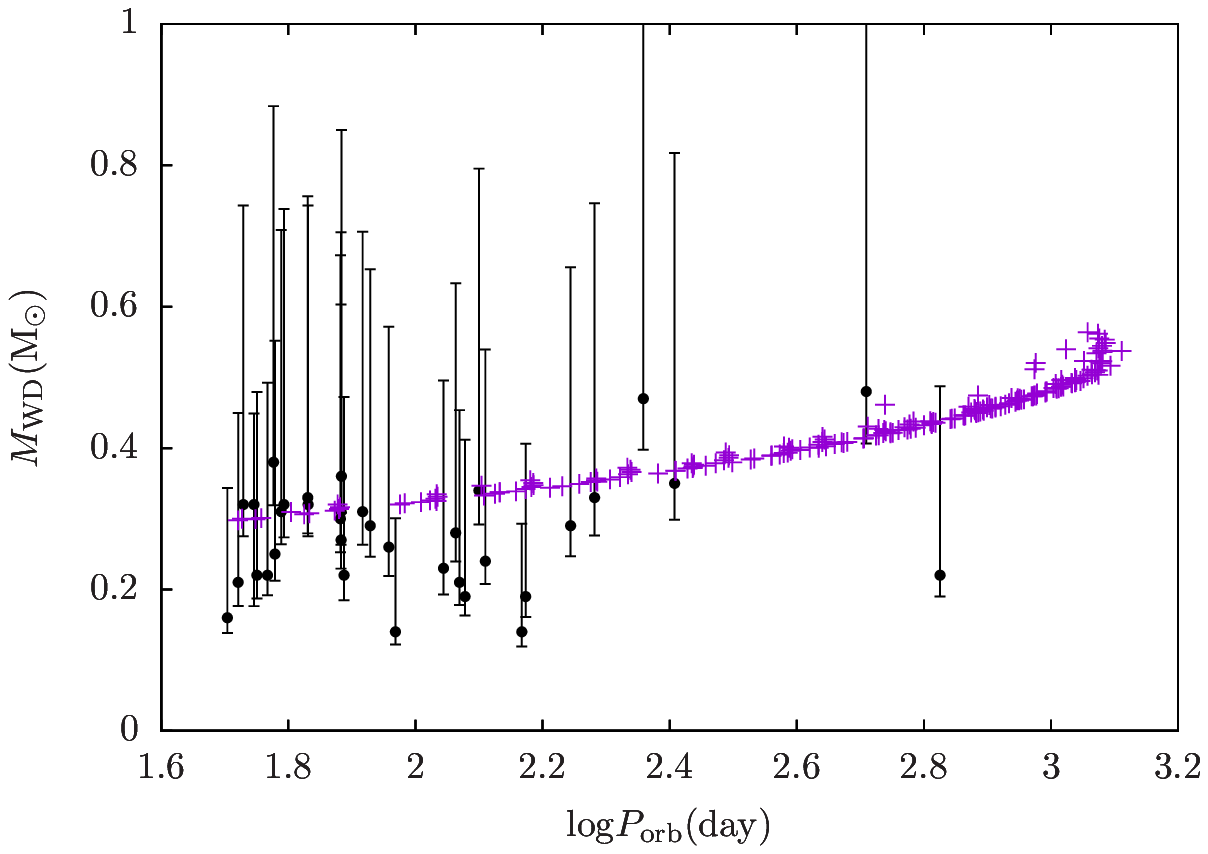,angle=0,width=11.5cm}
 \caption{Final WD companion masses of the MSPs formed via  the RG donor channel as a function of orbital periods.
The error bars represent the 35 observed MSPs with $P_{\rm spin}<40\,{\rm ms}$ listed 
in Table\,2 (see Manchester et al. 2005).
}
  \end{center}
\end{figure}

In Table 2, we listed some relevant parameters of 39 observed binary pulsars that 
have WD companions with long orbital periods ($>$50\,d),
in which 35 of them are MSPs that have spin periods $<40\,{\rm ms}$.
As shown in this Table, four peculiar binary pulsars with long orbital periods ($>$200\,d)  have spin periods $>40\,{\rm ms}$,   
i.e. PSRs $\rm J1711-4322$, $\rm J1822-0848$, $\rm J2016+1948$ and $\rm J2204+2700$.
It is hard for the present work to explain the origin of these four  binary pulsars; they may originate from 
isolated pulsars that have captured WDs in dense globular clusters (see, e.g. Verbunt \& Freire 2014) 
or other formation  ways of pulsars 
(e.g. core-collapse or electron-capture supernovae; see, e.g. van den Heuvel 2009).

Fig. 4 shows the resulting binary MSPs formed via the RG donor channel in the final ${M}_{\rm WD}-P_{\rm orb}$ diagram. 
The RG donor channel can produce  binary MSPs with wide orbital periods ranging from 50\,d to 1200\,d, 
and the WD companions have masses in the range of $0.30-0.55\,\rm M_{\odot}$.
As shown in this figure, almost all the observed MSPs with wide orbits can be covered 
by the RG donor channel in the ${M}_{\rm WD}-P_{\rm orb}$ diagram.
Thus, we stress that the  RG donor channel provides a viable way to explain the observed MSPs with wide separations.
By using a detailed binary population synthesis method, Wang (2018) recently estimated that 
the  Galactic rates of AIC events through the RG donor channel are in the range of
$\sim1-3\times10^{-5}\,\rm yr^{-1}$, and that the numbers of the resulting MSPs are  
$\sim1-4\times10^{5}$ in the Galaxy.

In Fig. 4, we also see that all the final MSPs follow 
the well-known  (${M}_{\rm WD}, P_{\rm orb}$)
relation between the companion mass and  the orbital period,
which can be understood by a relation between the mass of the giant's degenerate core  and its radius
(see, e.g. Refsdal \& Weigert 1971; 
Savonije 1987; Rappaport et al. 1995; Tauris \& Savonije 1999; Zhang et al. 2021).
During the mass-transfer stage, the RG radius ($R_2$) is approximately equal to its Roche-lobe radius ($R_{\rm L}$). 
Meanwhile, the $R_{\rm L}$ relates to the binary separation (i.e. the $P_{\rm orb}$). 
At the end of the mass-transfer stage, the final mass of the RG approximately equals to the mass of its degenerate core,
resulting in the correlation between the final companion mass and its orbital period.
This relation was confirmed in different kinds of observed peculiar binaries, such as binary MSPs, 
blue straggler binaries, sdB binaries, etc (see, e.g. Rappaport et al. 1995; 
Carney, Latham \& Laird 2005; Chen et al. 2013a; Gosnell et al. 2019). 
On the basis of this relation, one can get the mass of the WD companion once the orbital period is obtained.

\begin{figure}
\begin{center}
\epsfig{file=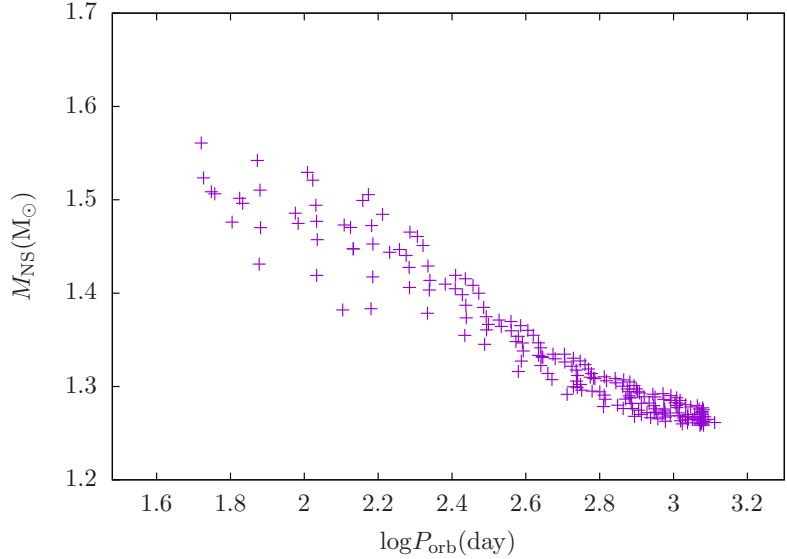,angle=0,width=11.5cm}
 \caption{Final NS masses of the MSPs formed via the RG donor channel as a function of orbital periods.}
  \end{center}
\end{figure}

Fig. 5 presents the resulting binary MSPs in the final ${M}_{\rm NS}-P_{\rm orb}$ diagram.
From this figure, we can see that there is an anti-correlation between the NS mass 
and the orbital period for the resulting MSPs,
i.e. MSPs with wide orbits have lower NS masses.
This is because systems with wide orbits experience higher mass-transfer process, 
losing too much material during pre-/post-AIC evolution.
Another reason is that systems with wide orbits have RG donors with larger degenerate cores, 
resulting in  less shell material being transferred onto the surface of NSs (see also Li et al. 2021).
The final NS masses in the binary MSPs   ranges from $1.26\,\rm M_{\odot}$ to $1.55\,\rm M_{\odot}$, in which
the accreted masses of the recycled pulsars for the RG donor channel are in the range of $\sim0.01-0.30\,\rm M_{\odot}$.

\begin{figure}
\begin{center}
\epsfig{file=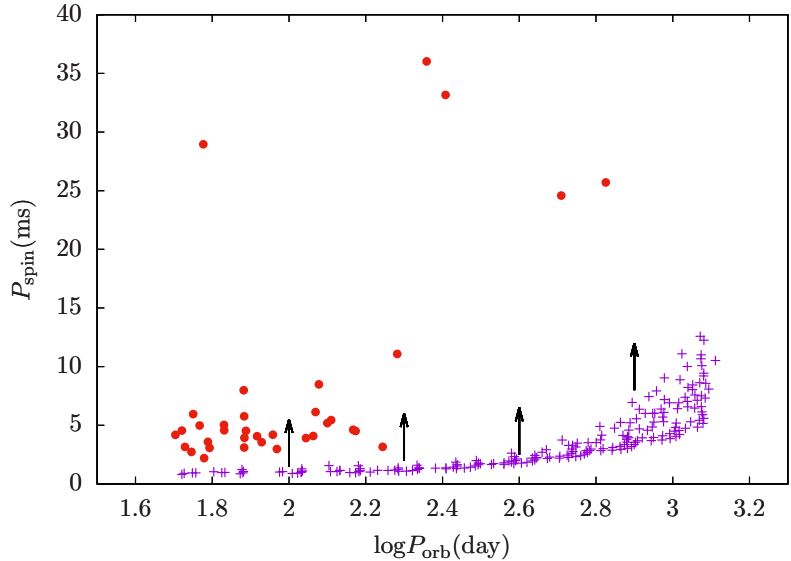,angle=0,width=11.5cm}
 \caption{Corbet diagram for the eventually formed MSPs based on the RG donor channel.
 The filled circles represent the 35 observed MSPs with $P_{\rm spin}<40\,{\rm ms}$ listed in Table\,2 (see Manchester et al. 2005).
The direction of the arrows means that the spin periods will become longer with the spin-down of the pulsars.}
  \end{center}
\end{figure}

Fig. 6 shows the Corbet diagram for the eventually formed MSPs based on the RG donor channel. 
The accreted mass of the recycled pulsars is up to $0.3\,\rm M_{\odot}$ (see Fig. 5), 
thus these MSPs will be fully recycled.
As shown in Fig. 6, the predicted minimum spin periods for different MSPs via the  RG donor channel 
are from 1\,ms to 10\,ms before the spin-down process.
With the spin-down of the pulsars, the spin periods will become longer.
This indicates that the present work has the potential ability to explain the observed MSPs with longer spin periods.
Note that there is a lack of MSPs with orbital periods $>$200\,d in observations. 
The main reason is as follows: (i) observational selection effect for long orbital separation systems, 
and (ii)  MSPs with wide orbits increase the possibility of disruption via stellar encounters, especially in globular clusters.

\section{Discussion}

\subsection{Comparison to previous studies}

Compared with previous studies (e.g. Li \& van den Heuvel 1997; Han \& Podsiadlowski 2004; Tauris et al. 2013), 
the present work enlarges the initial parameter regions of WD+RG systems forming WDs with $M_{\rm Ch}$,
especially the initial parameter region in this work containing more massive RG donors.
The main reason is that the present work adopted an adiabatic power-law assumptions for the mass-transfer  process,
which is more reasonable for semidetached binaries with RG donors (see also Liu et al. 2019).

Previous studies usually used a surface boundary criteria to compute the mass-transfer rate, written as
\begin{equation}
\dot{M}_{\rm 2}= -C\max[0,{(\frac{R_2}{R_{\rm L}}-1)^{\rm 3}}],
\end{equation}
in which $C$ is a dimensionless parameter usually set to be $\rm 1000\,\rm M_{\odot}\,yr^{\rm -1}$, 
$R_2$ is the radius of the donor, and $R_{\rm L}$ is the radius of its Roche-lobe (see Han, Tout \& Eggleton 2000). 
According to this assumption, the exceeding mass of the donor will be transferred onto the surface of  the accretor  
once RLOF occurs. 
Previous simulations indicate that the constant $C$ in Eq. (8) is too large for semidetached binaries with RG donors  during the mass-transfer process, 
probably overestimating the mass-transfer rate when the RG fills its Roche-lobe (see Liu et al. 2019, and references therein).
This leads to two cases that prevent the WD from growing in mass to $M_{\rm Ch}$, as follows:
(i) the WD loses too much mass in the form of the optically thick wind once the mass-transfer rate is larger than the critical rate $\dot{M}_{\rm cr}$, 
and (ii) a common envelope may be formed once the mass-transfer process becomes dynamically unstable.

In the present work, by supposing that the mass outflow is laminar and happens along the equipotential surface, 
and that the state equation of stars follow an adiabatic power law for RG donors, 
we used an approximate prescription for the mass-transfer process (see Eq. 1). 
In this case, the mass-transfer rate changes with the local material states, corresponding to a variable $C$ in Eq. (8).
Meanwhile, the mass-transfer rate for RG donors would be lower than that of previous studies, 
resulting in a lower mass-loss rate, and thus more material is retained onto the surface of the WD.
However, the integrated mass-transfer prescription for RG donors is still under debate. It has been
argued that a RG star  will not expand adiabatically once the timescale of the mass-loss is comparable with 
the local thermal timescale of the superadiabatic outer surface  (see Woods \& Ivanova 2011).
This indicates that the present work might underestimate the mass-transfer rate, 
but at least providing an upper limit of the initial parameter regions for producing binary MSPs.

We also note that Tauris et al. (2013) obtained a relatively small parameter space of ONe WD+RG systems forming WDs with $M_{\rm Ch}$.
This is mainly because Tauris et al. (2013) adopted a more strict model assumption for the evolution of ONe WD+RG systems; 
they supposed that a common envelope would be formed once the mass-transfer rate
is larger than three times of Eddington accretion rate of WDs.  
Tauris et al. (2013) emphasized that the critical criteria for the formation of  common envelope  
may be the largest uncertainties in their computations.

\subsection{Pre-AIC systems}

The pre-AIC systems with RG donors can be identified as the observed symbiotics that
play an important role for the evolution of semi-detached/detached binaries.
Symbiotics usually consist of a hot WD accretor and an evolved  RG companion with a wide orbit, in which 
the WD accumulates material from the RG donor via RLOF or the stellar wind  (see Truran \& Cameron 1971; 
Miko{\l}ajewska 2003; Podsiadlowski \& Mohamed 2007). 
Symbiotic novae are an important subgroup of symbiotics, where
the WD accretors undergo nova outbursts. Currently,
there are five symbiotic novae with WD masses 
close to ${M}_{\rm Ch}$ in observations, i.e. 
RS Oph (e.g. Hachisu \& Kato 2001; Brandi et al. 2009; Miko\l{}ajewska \& Shara 2017), 
T CrB (e.g. Belczy$\acute{\rm n}$ski \& Miko\l{}ajewska 1998; I{\l}kiewicz et al. 2016), 
V745 Sco (e.g. Drake et al. 2016; Orlando, Drake \& Miceli 2017),
V3890 Sgr (see  Miko\l{}ajewska et al. 2021)
and V407 Cyg (see I{\l}kiewicz et al. 2019).
However, it is still unclear 
whether the WD in these symbiotics is a CO WD or an ONe WD.

There is a main evolutionary route to produce ONe WD$+$RG systems that can 
form  AIC events and later evolve to recycled binary MSPs (for details see Fig. 3 of Wang \& Liu 2020).
MSPs in the RG donor channel originate from wider primordial binaries, in which  
the initial parameters of the primordial systems are in the range of $P^{\rm i}\sim1000-6000\,\rm d$,
$M_{\rm 1,i}\sim6-8\, M_{\odot}$ and $M_{\rm2,i}/M_{\rm1,i}<0.3$, where 
$P^{\rm i}$, $M_{\rm 1,i}$ and $M_{\rm 2,i}$  are the initial orbital period,  the initial masses of 
the primordial primary and secondary, respectively (see Wang \& Liu 2020).

\subsection{Post-AIC systems}

The post-AIC systems with RG donors can be identified to be
LMXBs in the observations.  Compared with the MS/He star donor channels,
the post-AIC systems with RG donors
will evolve into young MSPs with long orbital periods ($>$50\,d) eventually.
For the RG donor channel, the orbital separations at the moment of the AIC 
and the subsequent post-AIC systems (i.e. NS$+$RG systems) are relatively wide. 
In the observations,
there are two symbiotic X-ray binaries consisting of 
 accreting NSs and RG donors with known binary parameters (e.g. orbital periods and 
masses), as follows: (1) GX 1+4=V2116 Oph has a pulsar with a spin period of $\sim$2\,min 
and its orbital period is about 1161\,d, in which the pulsar
underwent one of the fastest spin-ups ever reported (see, e.g. Hinkle et al. 2006; I{\l}kiewicz, Miko{\l}ajewska \& Monard 2017).  
(2) 4U 1700+24=V934 Her hosts  a pulsar with an orbital period of $\sim$12\,yr, 
and the photospheric abundances with no trace of the NS-forming supernova event 
indicate that it is likely a post-AIC system (see Hinkle et al. 2019). 
In both systems, the RG donors are evolving towards the asymptotic giant branch, increasing the mass-transfer rate.

It is worth noting that such newborn NS systems with wide separations are thought to
increase the disruption possibility via stellar encounters in dense clusters,
possibly resulting in the formation of  isolated young NSs in globular clusters  
(see, e.g. Tauris et al. 2013; Verbunt \& Freire 2014; Belloni et al. 2020).
Freire \& Tauris (2014) recently studied the formation of MSPs 
from a rotationally delayed AIC of a super-${M}_{\rm Ch}$ WD, 
which can reproduce the observed eccentricities of  two MSPs in Galactic field, i.e. 
PSRs J2234$+$06 and J1946$+$3417.
For more studies on the formation of MSPs from LMXBs, see, e.g. 
Podsiadlowski, Rappaport \& Pfahl (2002), Nelemans \& Jonker (2010), Chen et al. (2013b, 2021), 
L\"{u} et al. (2017), Tauris (2018), Ablimit (2019),  Chen, Liu \& Wang (2020), Wang et al. (2021), etc.

\subsection{AIC events}

For the RG donor channel, it is expected to have a dense circumstellar material (CSM) surrounding the pre-AIC systems.
During the AIC process, the ejecta from the AIC collides with the dense CSM, probably producing a strong shock that
can form synchrotron radiation in radio frequencies.
It has been suggested that AIC events through the RG donor channel may 
show as radio bright but optically faint transients (see, e.g. Piro \& Thompson 2014; Moriya 2016).
Piro \& Thompson (2014) argued that  
the AIC ejecta collides with the RG companion, likely forming a  strong X-ray  radiation lasting for $\sim$1\,hr 
followed by an optical signal peaking at an absolute magnitude of $\sim-16$ to $-18$, where the optical signal lasts for a few days to a week.
It is expected that the radio signals of AIC can be detected by some radio transient surveys, such as 
the Square Kilometer Array transient survey and the Very Large Array Sky Survey   (see Moriya 2016).

There is no reported direct detection for AIC events so far, but there are some possible candidates of  
such events in observations (for a recent review see Wang \& Liu 2020), as follows:
(1) 
VTC J095517.5$+$690813 (a radio transient; see, e.g. Anderson et al. 2019; Moriya 2019).
(2) 
AT2018cow (one of the brightest fast-rising blue optical transients;  see, e.g. Tonry et al. 2018;
 Prentice et al. 2018; Margutti et al. 2019; Yu, Chen \& Li 2019).
(3) 
SN 2018kzr (one of the fastest declining supernova-like transients; see McBrien et al. 2019).
(4) 
The progenitor of fast radio burst FRB 121102 (a persistent radio source; see, e.g. Scholz et al. 2016; 
Margalit, Berger \& Metzger 2019; Waxman 2017; Sharon \& Kushnir 2020).
In order to confirm AIC events  and clarify the long-term issue  faced by the current theory of stellar evolution, 
more observational identifications and numerical simulations are needed.

\section{Summary}

In the present work, we study the formation of binary MSPs 
through the RG donor channel  systematically using an integrated mass-transfer prescription for RG donors. 
We found that 
the RG donor channel can form binary MSPs with orbital periods ranging from $50\,{\rm d}$ to  $1200\,{\rm d}$, in which
the final NS masses are in the range of  $\sim1.26-1.55\,\rm M_{\odot}$ 
and the masses of the WD companions are  in the range of $0.30-0.55\,\rm M_{\odot}$. 
We also found that the formed MSPs through the  RG donor channel  follow  
the correlation between the companion mass and  the orbital period, and that
there exists an anti-correlation between the final NS mass and the final orbital period. 
The pre-AIC systems with RG donors will show as symbiotics in the observations, whereas
the post-AIC systems can be identified as LMXBs, finally evolving into
young MSPs with wide orbits ($>$50\,d) compared with the MS/He star donor channels.
We emphasize that 
the RG donor channel provides a viable way to account for the observed MSPs with long orbital periods.
More theoretical and observational studies on MSPs with wide separations would be helpful 
for our understanding of this kind of pulsar systems. 

\section*{Acknowledgments}
We acknowledge the anonymous referee for valuable comments that help to improve the paper.
This study is partly supported by the NSFC (Nos 11873085, 12073071 and
11903075), 
the Youth Innovation Promotion Association of CAS (Nos 2018076 and 2021058),
the Western Light Youth Project of CAS,
the Yunnan Fundamental Research Projects (Nos 2019FJ001, 202001AS070029, 
202001AT070058, 202101AW070003 and 202101AW070047),
and the science research grants from the China Manned Space Project (Nos CMS-CSST-2021-B07/A10/A13).

\section*{Data availability}
The data of the numerical calculations in this  article can be made available on request by contacting BW.

\label{lastpage}
\end{document}